\documentclass[preprint,12pt]{elsarticle}
\usepackage{amsmath}
\usepackage{amsfonts}
\usepackage{amssymb}
\usepackage{graphicx}
\usepackage{bm}
\numberwithin{equation}{section}
\begin{document}
\begin{frontmatter}
\title{On the dynamics of a particle on a cone}
\author{K. Kowalski and J. Rembieli\'nski}
\address{Department of Theoretical Physics, University
of \L\'od\'z, ul.\ Pomorska 149/153, 90-236 \L\'od\'z,
Poland}
\begin{abstract}
A detailed study of the classical and quantum mechanics of a free particle
on a double cone and the particle bounded to its tip by the harmonic
oscillator potential is presented.
\end{abstract}
\end{frontmatter}
\section{Introduction}
The quantum mechanics on conical spaces has attracted much attention
not only in early investigations \cite{1,2} but also in recent
papers \cite{3,4,5,6,7}.  Among different motivations for the study
of a quantum particle on a cone we only recall the context of the
$2+1$ dimensional quantum gravity \cite{1}, cosmic strings \cite{8}
and defects in various media \cite{9}.  Nevertheless, to our best
knowledge, all investigations of quantum mechanics on a cone were
restricted to the case of the cone with a single nappe or
equivalently the plane with a deficit angle and the situation of a
double cone, i.e.\ two cones placed apex to apex, has not yet been
pursued.  This is all the more surprising since the case of a double
cone is much more simple.  In particular, in contrast to the
situation with a single nappe, we have no troublesome questions like
that concerning behavior of a free particle moving in a generator 
towards a cone tip after reaching it, and there is no need to
analyze self-adjoint extensions of symmetric operators representing
observables of the system.  In this work we perform a detailed
analysis of the classical and quantum mechanics on a double cone
involving the case of a free particle and harmonic oscillator on the
cone.  An interesting feature of the dynamics on the double cone discussed 
in this paper is the instability of the classical rectilinear motion on 
the meridian in the sense of Lyapunov and its implications on the quantum 
level.  The paper is organized as follows.  In Sec.\ II we study the
dynamics of the free classical particle on the cone.  Section III is
devoted to the quantum mechanics on the cone in the case of the free
motion.  The classical harmonic oscillator on the cone is discussed
in Sec.\ IV.  In Sec.\ V we investigate the quantization of the
harmonic oscillator on the cone.
\section{Classical mechanics of a free particle on a cone}
Let us consider a particle confined to a surface of a circular
double cone given by
\begin{eqnarray}
%<2.1>
x_1 &=& l\sin\alpha\cos\varphi,\nonumber\\
x_2 &=& l\sin\alpha\sin\varphi,\\
x_3 &=& l\cos\alpha,\nonumber
\end{eqnarray}
where $l\in(-\infty,\infty)$ is the coordinate of a particle on a
meridian (generator), $2\alpha$ is the opening angle of the cone, so
$\alpha\in(0,{\pi\over 2})$, and $\varphi\in[0,2\pi)$ specifies the
position of a particle on a parallel.  In view of (2.1) the equation
of the conical surface is of the form 
\begin{equation}
%<2.2>
x_1^2+x_2^2-{\rm tg}^2\alpha\, x_3^2 = 0.
\end{equation}
The Lagrangian of the free particle with mass $m$ constrained to the
conical surface (2.2) is
\begin{equation}
%<2.3>
L = \frac{m{\dot l}^2}{2} + \frac{m\sin^2\alpha l^2{\dot\varphi}^2}{2}.
\end{equation}
The corresponding Hamiltonian can be written as
\begin{equation}
%<2.4>
H = \frac{p_l^2}{2m} + \frac{p_\varphi^2}{2ml^2\sin^2\alpha},
\end{equation}
where $p_l=\frac{\partial L}{\partial{\dot l}}=m{\dot l}$ is the
momentum referring to the motion in the meridian and
$p_\varphi=\frac{\partial
L}{\partial{\dot\varphi}}=ml^2\sin^2\alpha{\dot\varphi}$ is the
angular momentum.  Therefore, the Hamilton's equations are
\begin{equation}
%<2.5>
\begin{split}
\dot l &= \frac{p_l}{m},\\
\dot \varphi &= \frac{p_\varphi}{ml^2\sin^2\alpha},\\
\dot p_l &= \frac{p_\varphi^2}{ml^3\sin^2\alpha},\\
\dot p_\varphi &= 0.
\end{split}
\end{equation}

Now let $J=p_\varphi={\rm const}$ designates the conserved angular
momentum.  For $J=0$ we find from (2.5) $\varphi={\rm const}$,
$p_l={\rm const}$, and
\begin{equation}
%<2.6>
l = l_0 + \frac{p_l}{m}t.
\end{equation}
Therefore $J=0$ refers to the rectilinear uniform motion along a
meridian.  

Consider now the case $J\ne0$.  Taking into account the following
expression on the energy 
\begin{equation}
%<2.7>
E = \frac{m{\dot l}^2}{2} + \frac{J^2}{2ml^2\sin^2\alpha},
\end{equation}
we find
\begin{equation}
%<2.8>
l = \pm\sqrt{\frac{2E}{m}(t+C)^2 + \frac{J^2}{2mE\sin^2\alpha}},
\end{equation}
where + ($-$) sign refers to the motion on the upper (lower) nappe of
the cone and $C$ is an integration constant which can be fixed with
the help of the initial data.  Thus, it turns out that for $J\ne0$
the particle resides definitely on the upper or lower nappe and
there is no communication between these two regions of the
configuration space.  This means, among others, that the solution
to (2.5) with $p_\varphi=J=0$ referring to rectilinear motion across
the tip of the cone is unstable in the Lyapunov sense.  Indeed, an 
arbitrary small perturbation of the initial condition $J=0$ leads to 
large deviations of corresponding solutions.  Such instability is 
illustrated in Fig.\ 1 (bottom-right figure).  On the other hand, 
an immediate consequence of (2.8) is the following 
inequality:
\begin{equation}
%<2.9>
|l|\ge \frac{|J|}{\sqrt{2mE}\sin\alpha},
\end{equation}
which means that for $J\ne0$ the geodesics have the lower (upper)
bound on the upper (lower) nappe.  We remark that the limit
$J\to0$ for (2.8) is
\begin{equation}
%<2.10>
\lim_{J\to0}l(t) = \Big|\frac{p_{l0}}{m}t+l_0\Big|,
\end{equation}
where we confined for brevity to the case of the motion on the upper 
nape. Clearly this asymptotics does not correspond to any real motion.
If so it would violate the uniqueness of the initial conditions
problem (Cauchy problem) related to the system (2.5).  An example of the
free motion on the cone (geodesic) given by (2.1) and (2.5) is presented
in Fig 1.  
\begin{figure*}
\centering
\begin{tabular}{c@{}c}
\includegraphics[width =.5\textwidth]{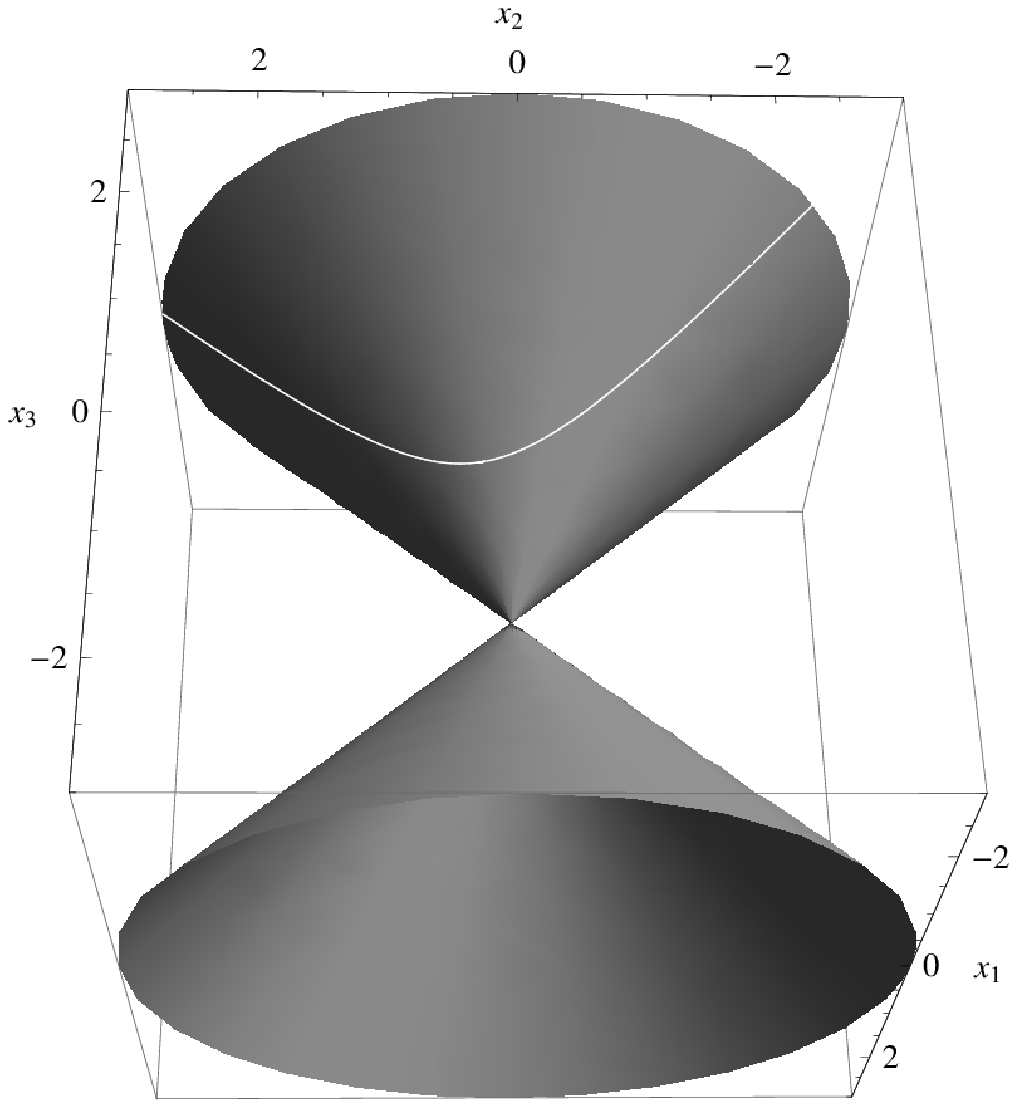}&
\includegraphics[width =.5\textwidth]{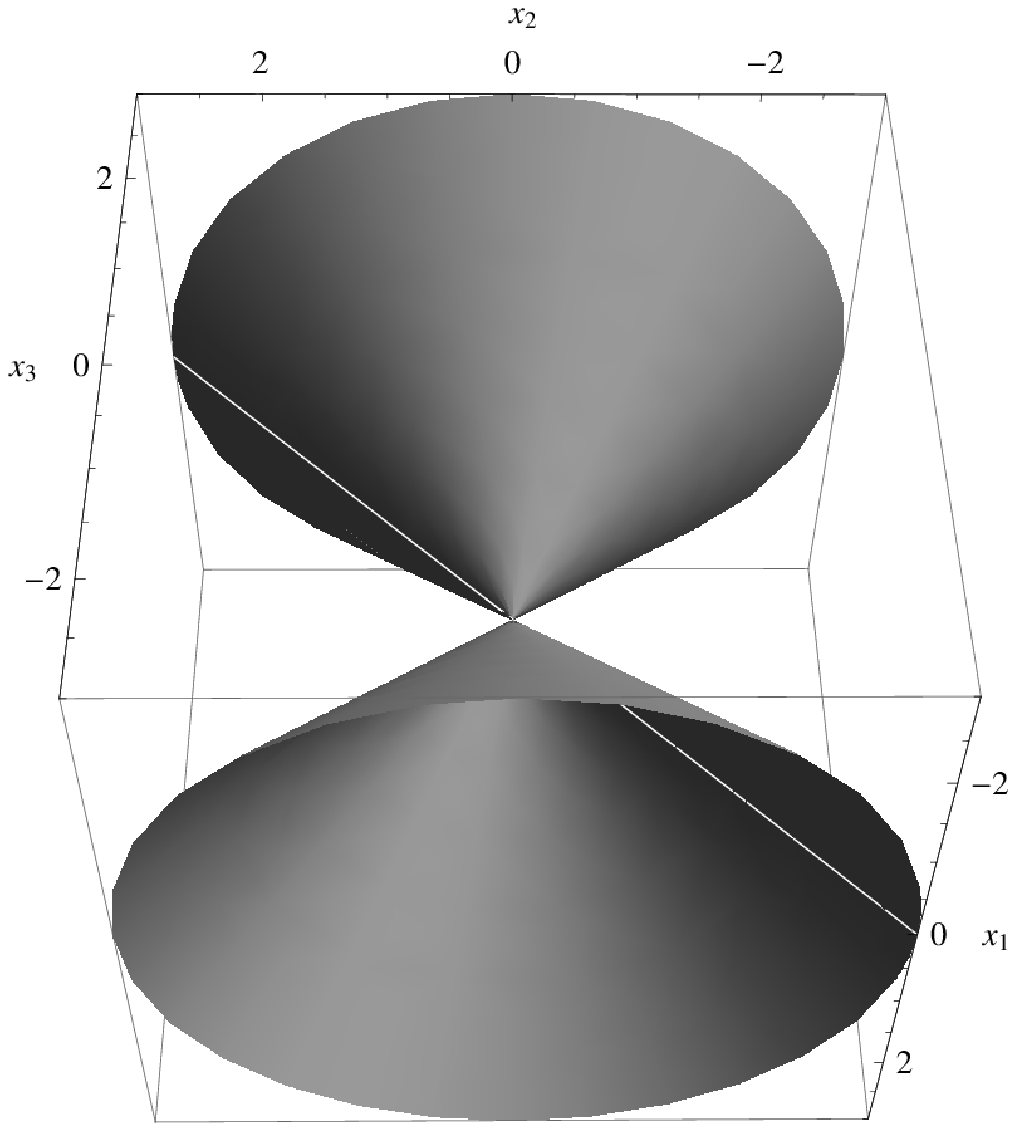}\\
\includegraphics[width =.5\textwidth]{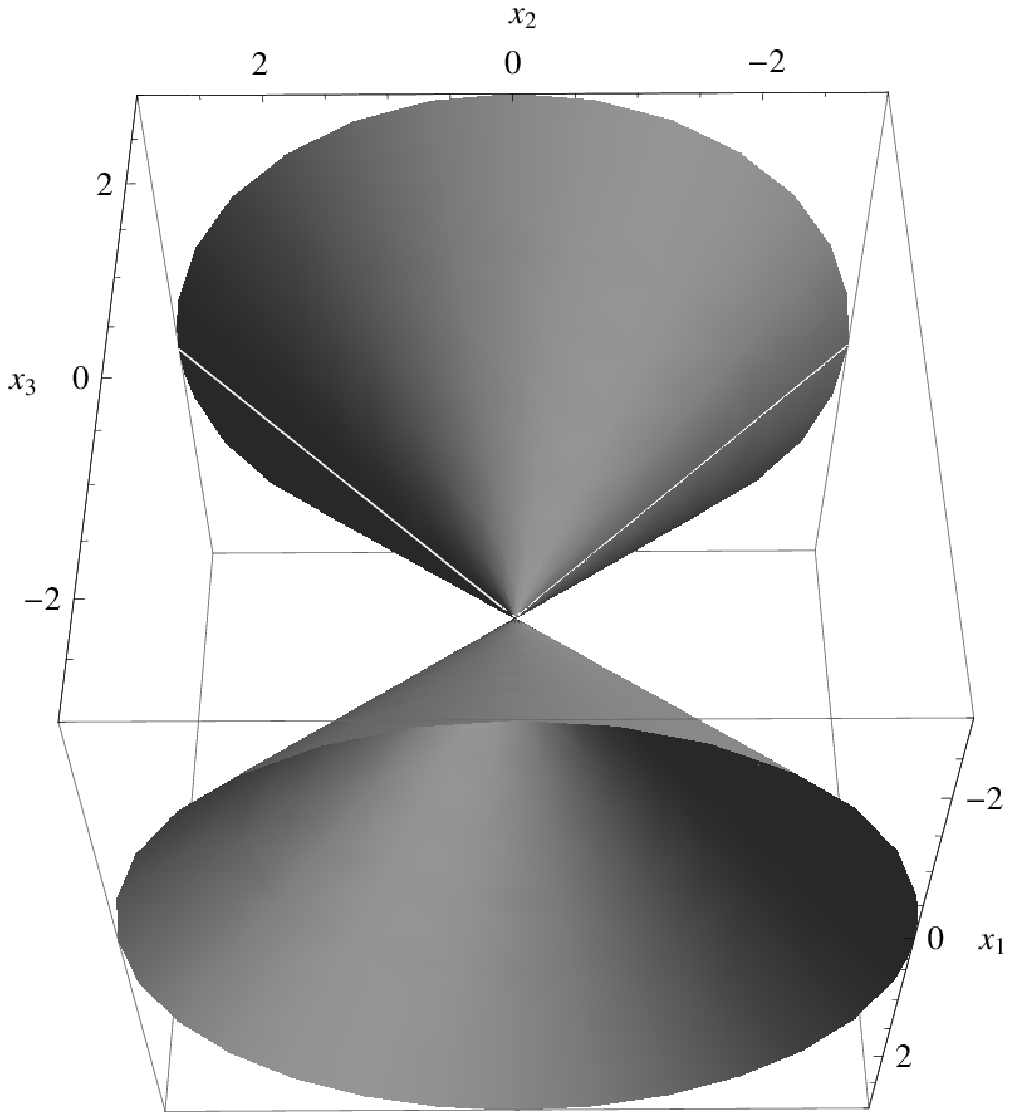}&
\includegraphics[width =.5\textwidth]{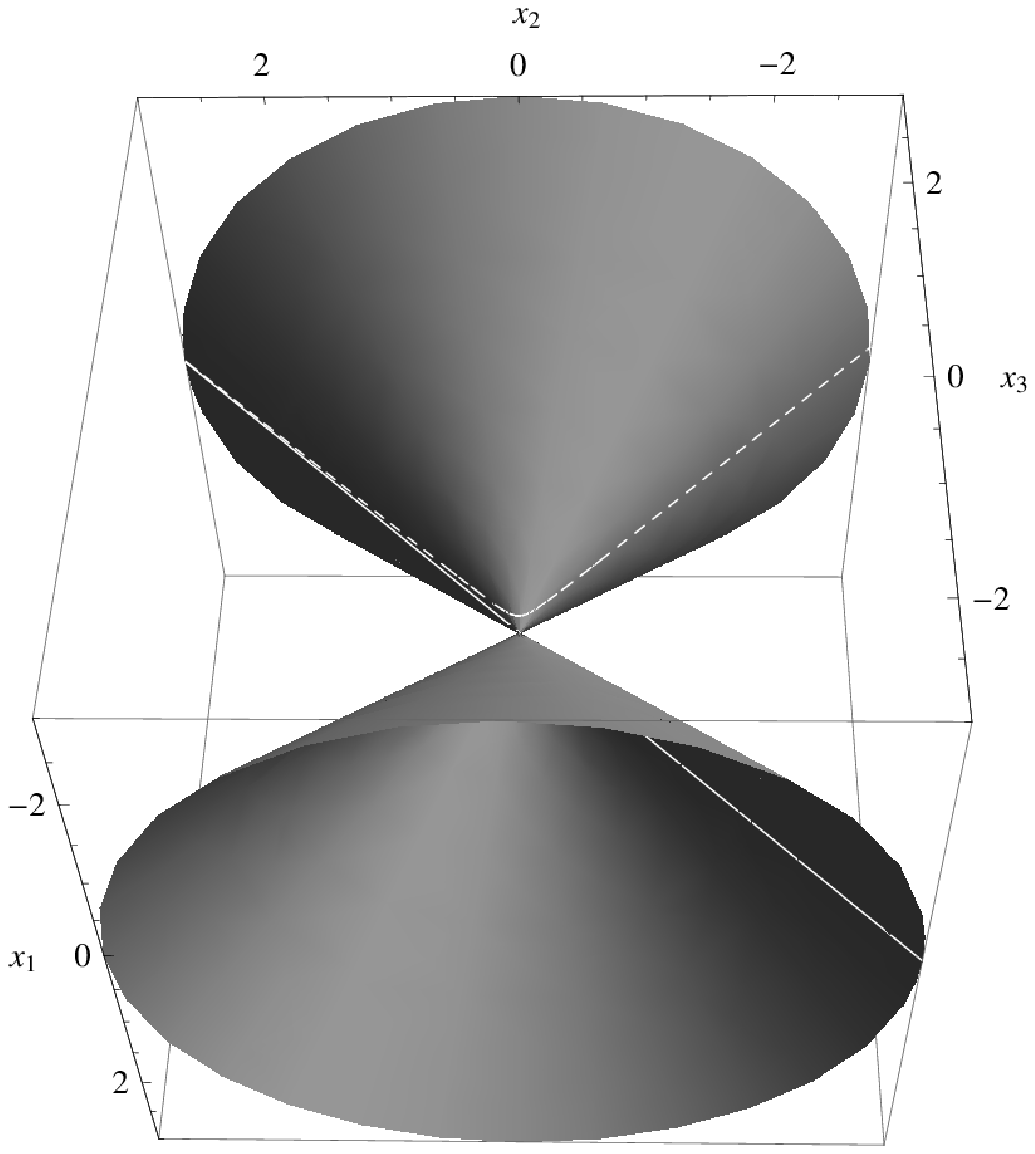}
\end{tabular}
\caption{Top left: the trajectory (geodesics) on the cone (2.1) which is 
the solution of (2.5) with the initial data $l_0=5$, $\varphi_0=\pi/2$, 
$p_{l0}=-1$, $J=1$, and the parameters $\alpha=\pi/4$ and $m=1$.  Top right:
the rectilinear motion in the meridian of the cone (geodesics) referring to
$J=0$ (see (2.6)).  The parameters and remaining initial conditions are the same
as in the figure on the left.  Bottom left: the trajectory given by (2.1) and 
(2.5) with the same parameters and initial data as in the figures above 
besides of $J=0.01$ corresponding to the limit (2.10) and (2.13). Bottom right:
solid line: the solution to (2.5) with $J=0$ presented in top-right figure.  
Dashed line: the solution to (2.5) with the same parameters and initial 
conditions as in the top-left figure besides $J=0.1$.}
\end{figure*}
Further, from the second equation of (2.5) we get
\begin{equation}
%<2.11>
\varphi-\varphi_0 = 2E\sin\alpha\left[{\rm
arctg}\frac{2E\sin\alpha}{J}(t+C)-{\rm
arctg}\frac{2E\sin\alpha}{J}C\right],\qquad {\rm for}\quad J\ne0,
\end{equation}
and
\begin{equation}
%<2.12>
\varphi={\rm const},\qquad {\rm for}\quad J=0.  
\end{equation}
The limit $J\to0$ for the angle (2.11) is
\begin{equation}
%<2.13>
\varphi-\varphi_0=0\quad {\rm or}\quad |\varphi-\varphi_0|=\pi.
\end{equation}
Finally from the first equation of (2.5) and (2.8) we find
\begin{equation}
%<2.14>
p_l =
\pm\frac{2E(t+C)}{\sqrt{\frac{2E}{m}(t+C)^2+\frac{J^2}{2mE\sin^2\alpha
}}},\qquad {\rm for}\quad J\ne0.
\end{equation}
By means of (2.8) $p_l$ vanishes when the minimal value is reached
by $|l|$ specified by (2.9).  The limit $J\to0$ for the momentum in
the case of the upper nappe can be written as
\begin{equation}
%<2.15>
\lim_{J\to0} p_l(t) =
|p_{l0}|\varepsilon\left(t+\frac{ml_0}{p_{l0}}\right),
\end{equation}
where $l_0>0$ and $\varepsilon(x)$ is the sign function.
\section{Quantum mechanics of a free particle on a cone}
In quantum mechanics the dynamics of a free particle on a double cone
defined by the Hamiltonian (2.4), is described by the Schr\"odinger
equation
\begin{equation}
%<3.1>
{\rm i}\frac{\partial f(l,\varphi;t)}{\partial t}=\left(\frac{{\hat
p_l}^2}{2m} + \frac{{\hat J}^2}{2m{\hat
l}^2\sin^2\alpha}\right)f(l,\varphi;t),
\end{equation}
where $\hat p_l$, $\hat l$, and $\hat J$ are the momentum, position
and angular momentum operators, respectively, and we set $\hbar=1$.  
The Hilbert space for a quantum particle on a cone is specified by 
the scalar product
\begin{equation}
%<3.2>
\langle f|g\rangle =
\int_0^{2\pi}d\varphi\int_{-\infty}^{\infty}dl\,|l|f^*(l,\varphi)g(l,\varphi),
\end{equation}
where $|l|dld\varphi$ coincides up to the multiplicative constant
with the surface element of the cone defined by (2.1) such that
$dS=\sin\alpha|l|dld\varphi$.  The operators $\hat p_l$, $\hat l$
and $\hat J$ act in the representation (3.2) in the following way:
\begin{eqnarray}
%<3.3>
{\hat p_l}f(l,\varphi) &=& -{\rm i}\left(\frac{\partial}{\partial l}
+\frac{1}{2l}\right)f(l,\varphi),\nonumber\\
{\hat l}f(l,\varphi) &=& lf(l,\varphi),\\
{\hat J}f(l,\varphi) &=& -{\rm
i}\frac{\partial}{\partial\varphi}f(l,\varphi)\nonumber.
\end{eqnarray}
We remark that the action of the operator ${\hat p_l}$ different
from the Schr\"odinger representation ensures its hermicity with
respect to the scalar product (3.2).  Of course, such form of the
operator ${\hat p_l}$ preserves the canonical commutation relations
of ${\hat p_l}$ and ${\hat l}$ as well.  The counterpart of the operator
$\hat p_l$ in the case of the cone with a single nappe was introduced
in reference 4.  Nevertheless, it is not self-adjoint so it cannot be
regarded as a physical observable.  We stress that in opposition to the
case of the single nappe (or the plane with deficit angle) where the
observables such as for example energy are labelled by parameters 
related to their self-adjoint extensions \cite{4,5}, the operators 
representing physical observables for the quantum mechanics on the 
double cone discussed herein, are defined uniquely.  We finally point
out that the analysis of self-adjoint extensions in the case of the
cone with a single nappe is closely related to the study of self-adjoint 
extensions describing a quantum particle on a plane with extracted point 
performed by us in reference 10.

Consider now the eigenvalue equation for the Hamiltonian ${\hat
H}f(l,\varphi)~=~Ef(l,\varphi)$ following directly from (3.1) and (3.3)
\begin{equation}
%<3.4>
\frac{1}{2m}\left(-\frac{\partial^2}{\partial
l^2}-\frac{1}{l}\frac{\partial}{\partial l} + \frac{1}{4l^2} -
\frac{1}{l^2\sin^2\alpha}\frac{\partial^2}{\partial\varphi^2}\right)f_E(l,\varphi)
=Ef_E(l,\varphi).
\end{equation}
On separating variables we find
\begin{equation}
%<3.5>
f_{j,E}(l,\varphi)=e^{{\rm i}j\varphi}u_{j,E}(l),
\end{equation}
where $f_{j,E}(l,\varphi)$ are the common eigenvectors of the Hamiltonian and the
operator of the angular momentum $\hat J$, and $u_{j,E}(l)$ satisfies the equation
\begin{equation}
%<3.6>
\frac{d^2u_{j,E}(l)}{dl^2}+\frac{1}{l}\frac{du_{j,E}(l)}{dl}+\left(2mE-
\frac{\frac{1}{4}+\frac{j^2}{\sin^2\alpha}}{l^2}\right)u_{j,E}(l)=0.
\end{equation}
The solution to (3.6) for $j\ne0$ with convergent norm (in a
distributive sense) is of the form
\begin{equation}
%<3.7>
u_{j,E}(l) =
CJ_{\sqrt{\frac{1}{4}+\frac{j^2}{\sin^2\alpha}}}(\sqrt{2mE}l),
\end{equation}
where $J_\nu(z)$ is the Bessel function of the first kind and $C$ is a
normalization constant.  Now, bearing in mind the behavior of the
classical free particle for $J\ne0$ discussed in previous section,
taking into account that $J_\nu(0)=0$ for $\nu>0$, and the identity
\begin{equation}
%<3.8>
J_\nu(-z)=e^{\nu\pi{\rm i}}J_\nu(z),
\end{equation}
it seems plausible to identify the case $l>0$ ($l<0$) in (3.7) with
the quantum particle on the upper (lower) nappe.  Let us designate
by $u_{+j,E}(l,\varphi)$ ($u_{-j,E}(l,\varphi)$) the corresponding
solution to (3.6), so
\begin{eqnarray}
%<3.9>
u_{+j,E}(l)&=&\begin{cases}J_{\sqrt{\frac{1}{4}+\frac{j^2}{\sin^2\alpha}}}(\sqrt{2mE}l),
&\mbox{for } l>0,\\0,&\mbox{for } l\le0,\end{cases}\\
u_{-j,E}(l)&=&\begin{cases}0,&\mbox{for } l\ge0,\\J_{\sqrt{\frac{1}{4}+\frac{j^2}{\sin^2\alpha}}}
(\sqrt{2mE}l), &\mbox{for } l<0.\end{cases}
\end{eqnarray}
In general case the state of a particle on a cone is the
superposition of the states $u_{+j,E}$ and $u_{-j,E}$, so we finally
arrive at the solution to (3.6) such that
\begin{equation}
%<3.11>
f_{j,E}(l,\varphi) = e^{{\rm
i}j\varphi}[Au_{+j,E}(l)+Bu_{-j,E}(l)],\qquad j\ne0,
\end{equation}
where $A$ and $B$ are constant.  Taking into account the so called
``closure equation'' \cite{11} of the form
\begin{equation}
%<3.12>
\int_0^\infty xJ_\nu(\alpha x)J_\nu(\beta x)dx = \frac{1}{\alpha
}\delta(\alpha -\beta),\qquad \nu>-\frac{1}{2},
\end{equation}
and (3.8) we find for the eigenvectors of the Hamiltonian (3.11)
normalized as
\begin{equation}
%<3.13>
\langle f_{j,E}|f_{j'E'}\rangle = \int_0^{2\pi}d\varphi\int_{-\infty}^{\infty}dl\,|l|
f_{j,E}^*(l,\varphi)f_{j',E'}(l,\varphi)=\delta_{jj'}\delta(\sqrt{E}-\sqrt{E'}),
\end{equation}
the relation
\begin{equation}
%<3.14>
|A|^2 + |B|^2=\frac{m\sqrt{E}}{\pi}.
\end{equation}

Now, it can be easily checked that for $j=0$ the eigenvectors of
the Hamiltonian satisfying (3.6) are given by
\begin{equation}
%<3.15>
f_{0,E}(l,\varphi) =
\frac{1}{\sqrt{|l|}}[A\sin(\sqrt{2mE}l)+B\cos(\sqrt{2mE}l)].
\end{equation}
For $E^2+E'^2\ne0$ the coefficients $A$ and $B$ of the normalized
state (3.15) satisfy the condition $|A|^2+|B|^2=\sqrt{2m}/2\pi^2$.  
Otherwise, if $E=E'=0$ we have $|B|=\frac{1}{2\pi}$.  Taking into 
accout the identity
\begin{equation}
%<3.16>
J_{\frac{1}{2}}(z) = \sqrt{\frac{2}{\pi z}}\sin z,
\end{equation}
we find that in the limit $j\to0$ of (3.11) only a part of the
solution (3.15) referring to the rectilinear motion in the meridian
is obtained.  The lacking part could be derived as the limit $j\to0$
of the function $J_{-\sqrt{\frac{1}{4}+\frac{j^2}{\sin^2\alpha}}}
(\sqrt{2mE}l)$ by means of the identity
\begin{equation}
%<3.17>
J_{-\frac{1}{2}}(z) = \sqrt{\frac{2}{\pi z}}\cos z.
\end{equation}
However, the Bessel function $J_{-\sqrt{\frac{1}{4}+\frac{j^2}{\sin^2\alpha}}}
(\sqrt{2mE}l)$ has the index lesser than $-\frac{1}{2}$ and the
``closure equation'' (3.12) related to normalization cannot be
applied.  Such behavior of solutions (3.11) and (3.15) is a quantum
scar of the instability of the classical rectilinear motion in the
cone discussed in Sec.\ II.
\section{Classical mechanics of the harmonic oscillator on a cone}
We now discuss the motion of the classical particle on the double cone
bound to its tip by harmonic oscillator potential.  The coresponding
Hamiltonian is (see (2.4))
\begin{equation}
%<4.1>
H = \frac{p_l^2}{2m} + \frac{p_\varphi^2}{2ml^2\sin^2\alpha}
+ \frac{m\omega^2}{2}l^2,
\end{equation}
and the Hamilton's equations are of the form
\begin{equation}
%<4.2>
\begin{split}
\dot l &= \frac{p_l}{m},\\
\dot \varphi &= \frac{p_\varphi}{ml^2\sin^2\alpha},\\
\dot p_l &= \frac{p_\varphi^2}{ml^3\sin^2\alpha}-m\omega^2l,\\
\dot p_\varphi &= 0.
\end{split}
\end{equation}
Of course $J=p_\varphi=0$ which leads to $\varphi={\rm const}$,
refers to the case of the standard harmonic oscillator on a line
(generator) with the solution
\begin{equation}
%<4.3>
\begin{split}
l &= l_0\cos\omega t + \frac{p_{l0}}{m\omega}\sin\omega t,\\
p_l &= p_{l0}\cos\omega t - \omega ml_0\sin\omega t.
\end{split}
\end{equation}

Now, using the expression on the energy such that
\begin{equation}
%<4.4>
E = \frac{m{\dot l}^2}{2} + \frac{J^2}{2ml^2\sin^2\alpha}
+ \frac{m\omega^2}{2}l^2,
\end{equation}
we find for $J\ne0$
\begin{equation}
%<4.5>
l =
\pm\sqrt{\frac{E+\sqrt{E^2-\frac{J^2\omega^2}{\sin^2\alpha}}\sin2\omega
(t+C)}{m\omega^2}},\qquad J\ne0,
\end{equation}
where plus and minus sign correspond to the upper and lower nappe,
respectively, and $C$ is an integration constant.  It thus appears 
that as with the case of the free motion, whenever $J\ne0$ the particle 
remains in one concrete nappe during the time evolution.  Therefore, 
the solution (4.3) describing the harmonic oscillations in a meridian 
around the point $l=0$ turns out to be unstable (see Fig.\ 3).  On the 
other hand, an immediate consequence of (4.5) is that for $J\ne0$ the 
motion is finite.  Namely, we find
\begin{equation}
%<4.6>
\sqrt{\frac{E-\sqrt{E^2-\frac{J^2\omega^2}{\sin^2\alpha}}}{m\omega^2}}
\le |l| \le \sqrt{\frac{E+\sqrt{E^2-\frac{J^2\omega^2}{\sin^2\alpha}}}{m\omega^2}}.
\end{equation}
We remark that in the limit $J\to0$ we obtain the correct amplitude
$\sqrt{\frac{2E}{m\omega^2}}$ of the harmonic oscillations (4.3),
however the period of harmonic oscillations (4.3) with $J=0$ is
twice the period of oscillations (4.5) with $J\ne0$.  For the upper
nappe the limit $J\to0$ of (4.5) is
\begin{equation}
%<4.7>
\lim_{J\to0}l(t) =
\sqrt{\frac{E}{m\omega^2}}\sqrt{1+\sqrt{\frac{l_0^2\omega^2p_{l0}^2}{E^2}}\sin2
\omega t+\left(\frac{l_0^2m\omega^2}{E}-1\right)\cos2\omega t}.
\end{equation}
The trajectories on the cone (2.1) satisfying (4.2) involving the case
of small $J$ are shown in Fig 2.
\begin{figure*}
\centering
\begin{tabular}{c@{}c}
\includegraphics[width =.5\textwidth]{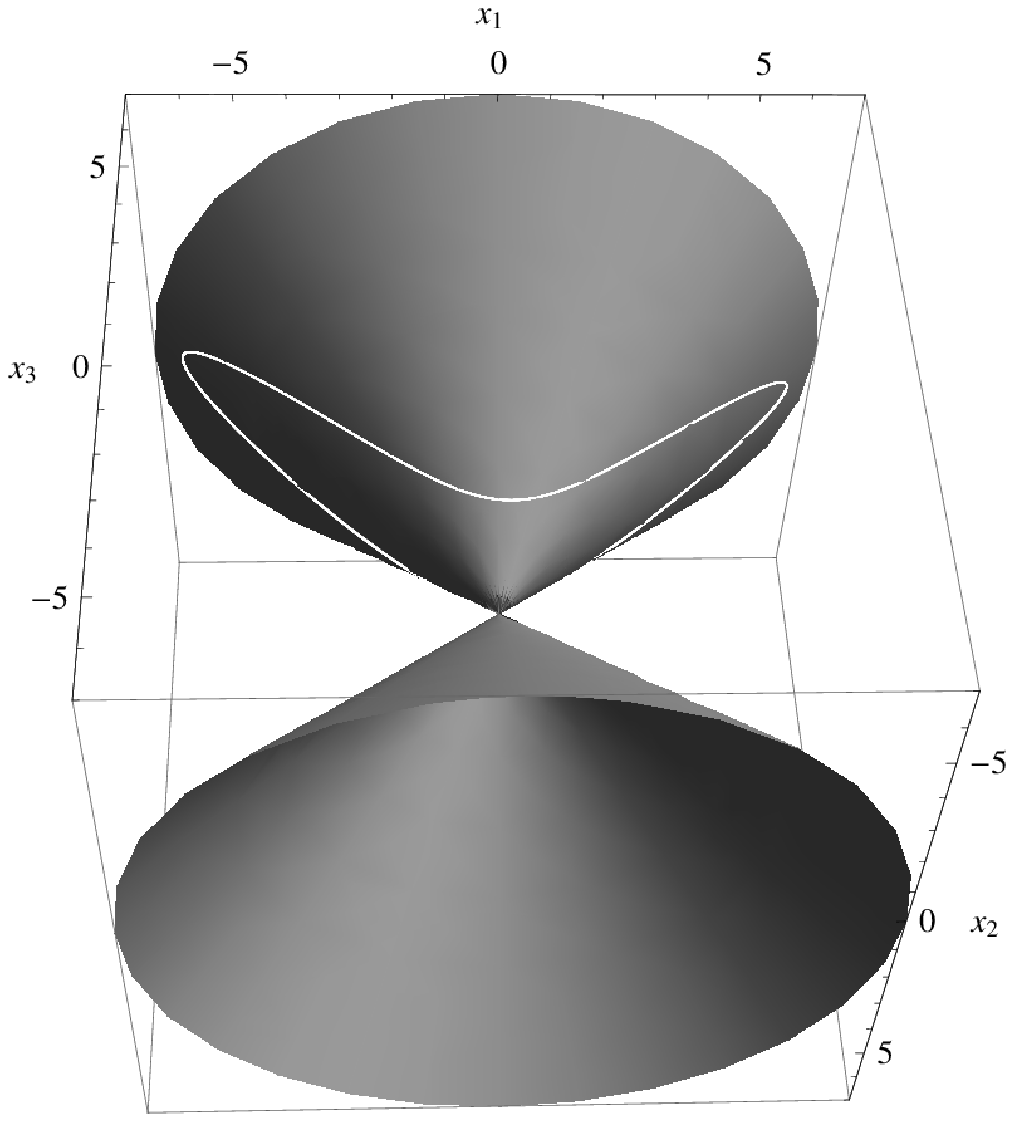}&
\includegraphics[width =.5\textwidth]{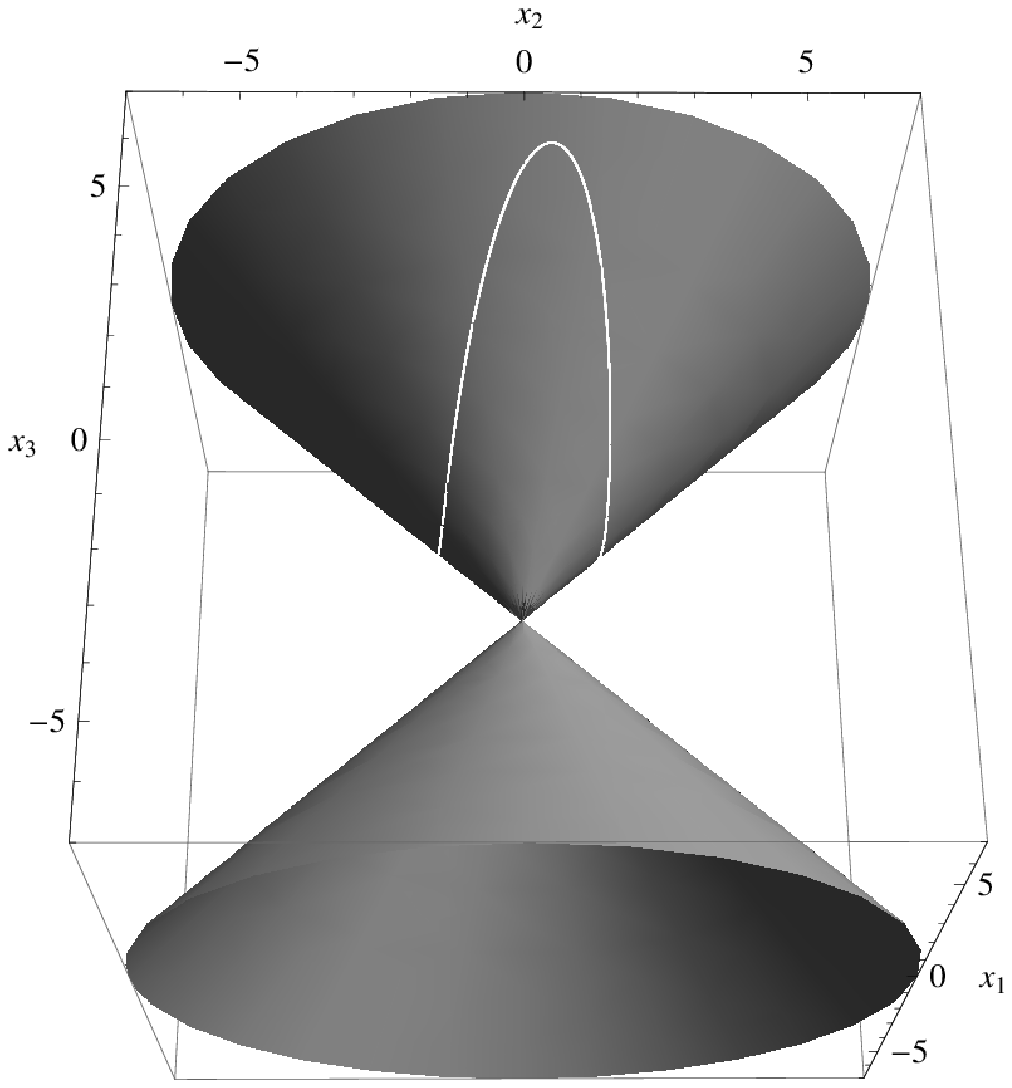}\\
\includegraphics[width =.5\textwidth]{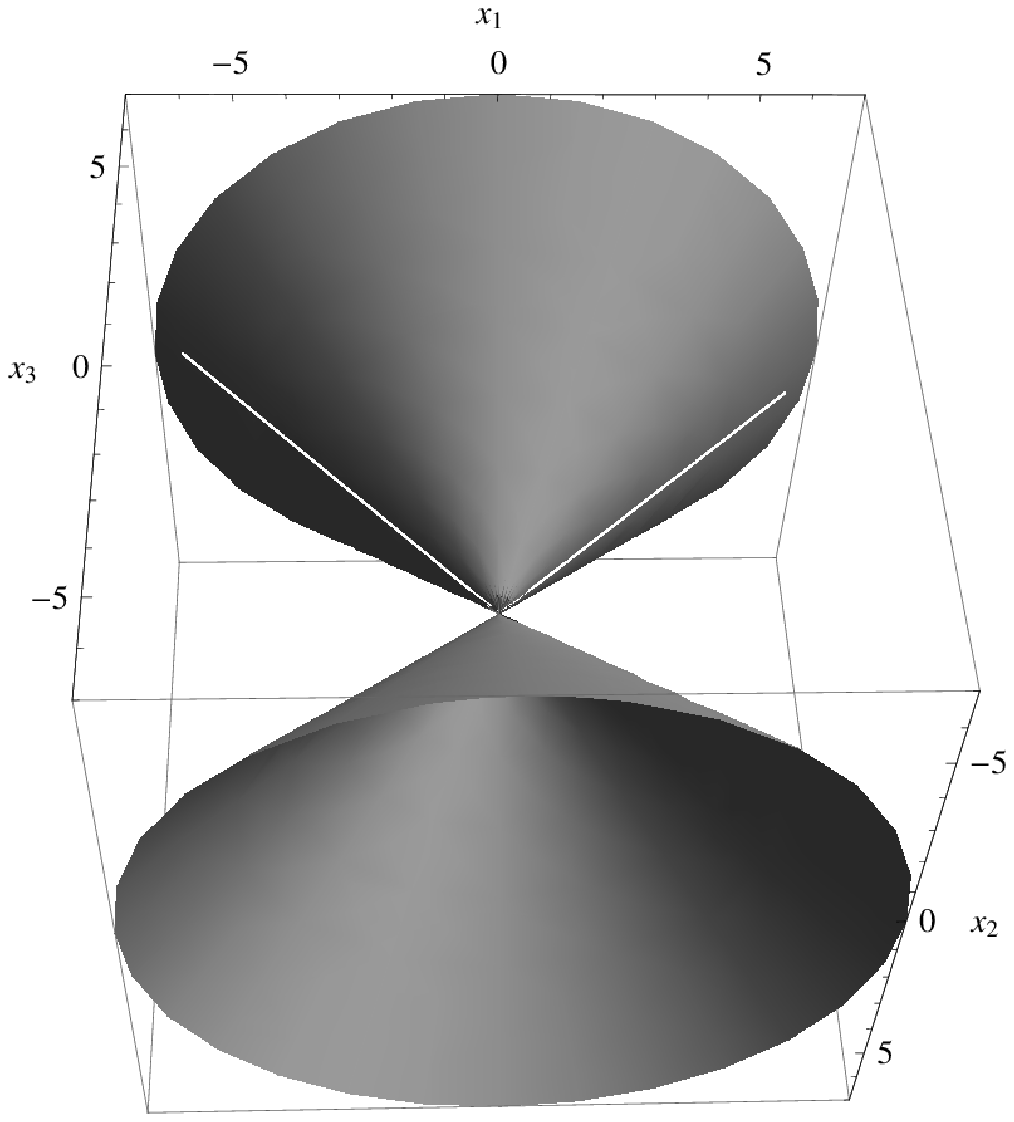}&
\includegraphics[width =.5\textwidth]{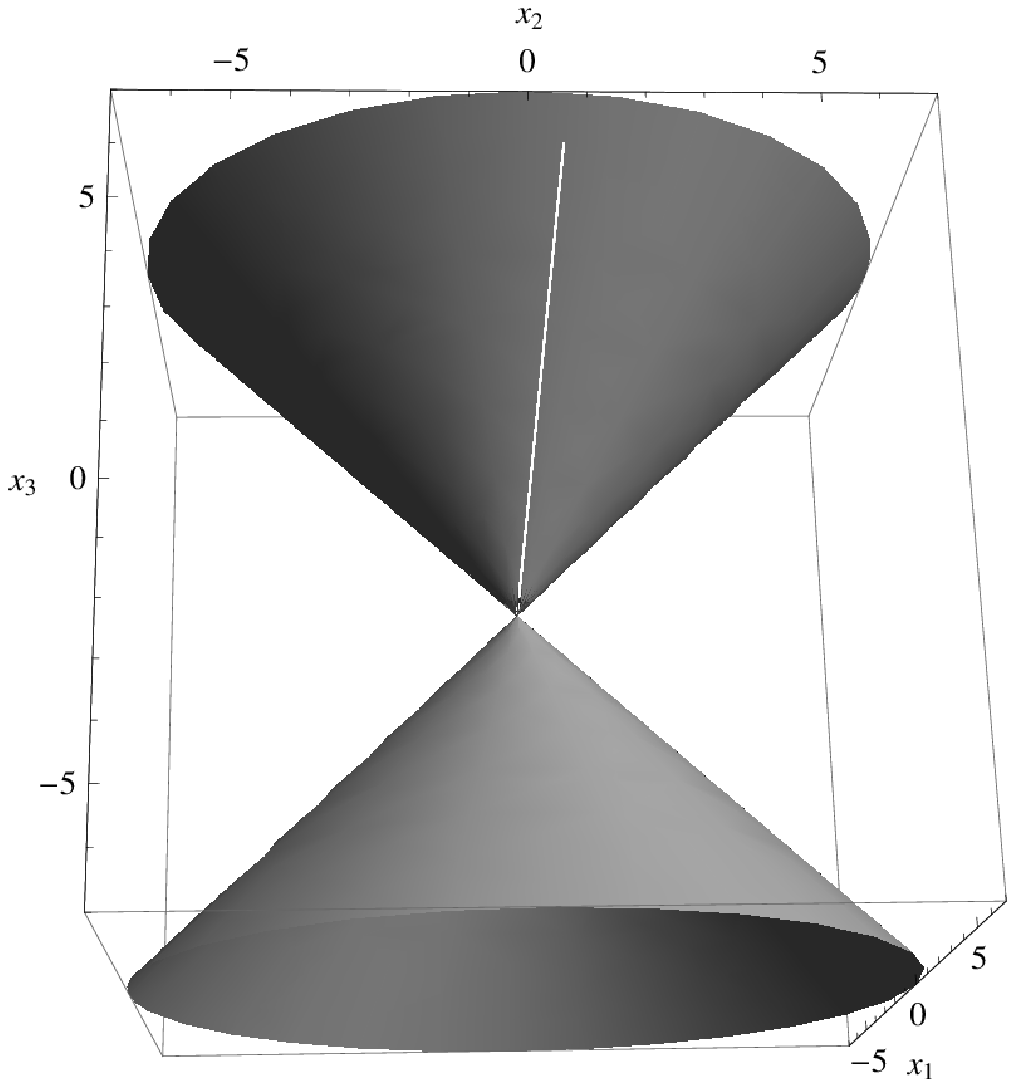}
\end{tabular}
\caption{Top left: the trajectory of the harmonic oscillator on the cone 
given by (2.1) and (4.2) subject to the initial data $l_0=9$, $\varphi_0=0.1$, 
$p_{l0}=-1$, $J=20$, and the parameters $\alpha=\pi/4$, $m=1$ and 
$\omega=\sqrt{2}$. View from below.  Top right: view of the right side of 
the trajectory on the cone presented in the figure on the left.  Bottom 
left: the motion of the harmonic oscillator on the cone with the parameters
and initial conditions the same as in the top figures besides of $J=0.01$
referring to the limit (4.7) and (4.9).  View from below.  Bottom 
right: right-hand view of the bottom-left figure.}
\end{figure*}
\begin{figure*}
\centering
\begin{tabular}{c@{}c}
\includegraphics[width =.5\textwidth]{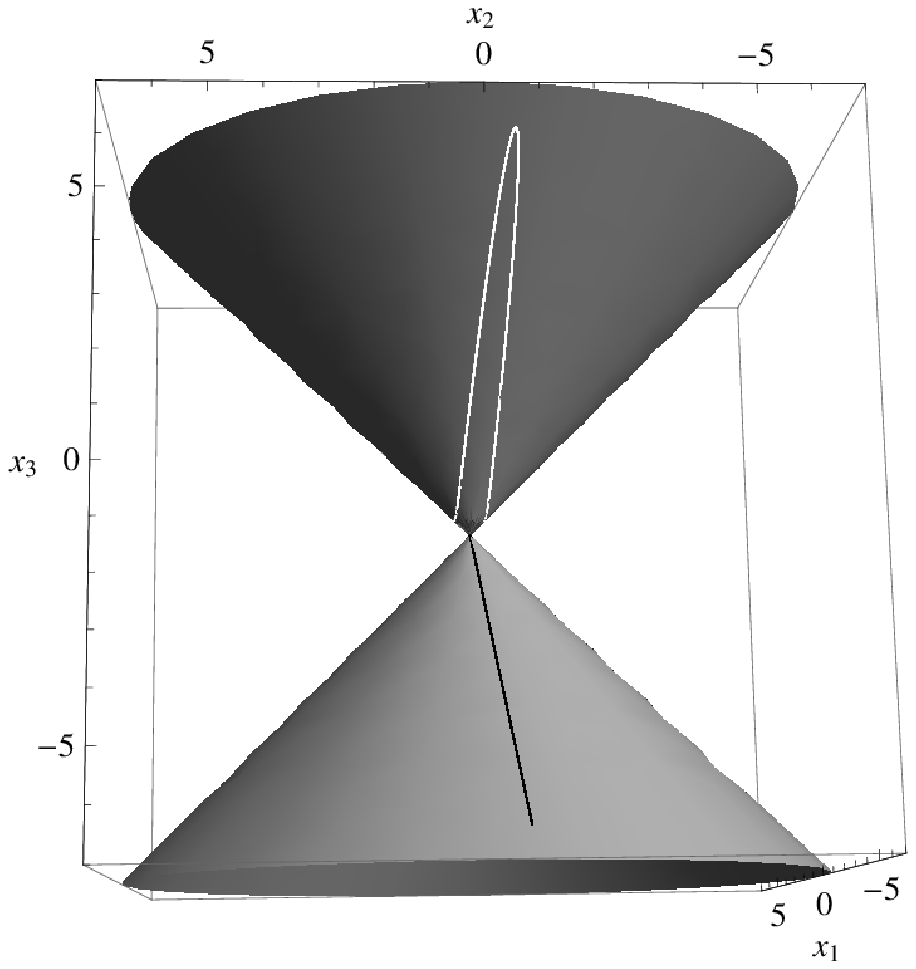}&
\includegraphics[width =.5\textwidth]{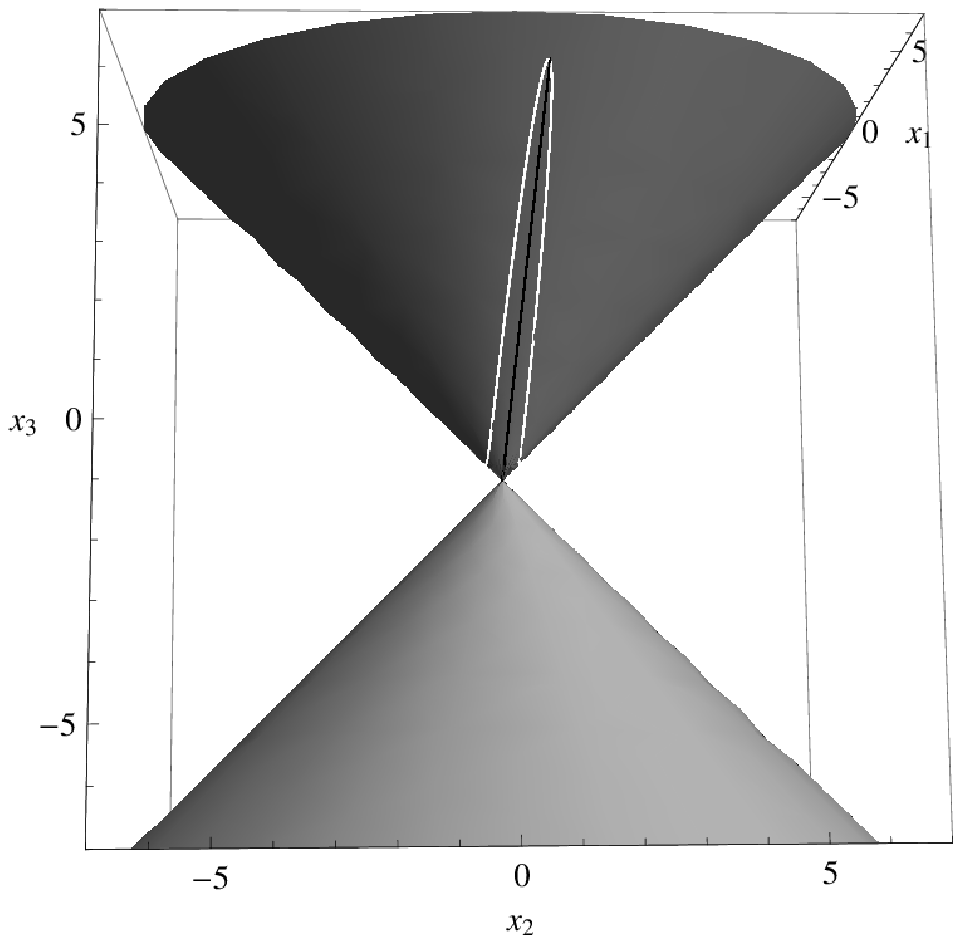}
\end{tabular}
\caption{The illustration of the instability of the harmonic oscillations 
on the cone corresponding to the vanishing angular momentum.  Left: white
line refers to the solution to (4.2) with the same values of parameters and
initial conditions as in the top figures in Fig.\ 2 besides of $J=4$.  Black
line is the part of the segment from the lower nappe which is the trajectory 
of the harmonic oscillator with $J=0$ (the remaining parameters and initial
data coincide with those for the white trajectory).  Right: the view of the
figure on the left from the behind.}
\end{figure*}

Furthermore, using the second equation of (4.2) we get for $J\ne0$
\begin{equation}
%<4.8>
\varphi-\varphi_0 = \frac{\varepsilon(J)}{\sin\alpha}\left({\rm arctg}
\frac{E{\rm tg}\omega(t+C)+\sqrt{E^2-\frac{J^2\omega^2}{\sin^2\alpha}}}
{\sqrt{\frac{J^2\omega^2}{\sin^2\alpha}}}
- {\rm arctg}
\frac{E{\rm tg}\omega C+\sqrt{E^2-\frac{J^2\omega^2}{\sin^2\alpha}}}
{\sqrt{\frac{J^2\omega^2}{\sin^2\alpha}}}\right).
\end{equation}
As with the case of the free motion we obtain in the limit $J\to0$
\begin{equation}
%<4.9>
\varphi-\varphi_0=0\quad {\rm or}\quad |\varphi-\varphi_0|=\pi.
\end{equation}
Finally, using the first equation of (4.2) we derive the following
formula on the momentum
\begin{equation}
%<4.10>
p_l =
\pm\frac{\sqrt{m}\sqrt{E^2-\frac{J^2\omega^2}{\sin^2\alpha}}\cos2\omega
(t+C)}{\sqrt{E+\sqrt{E^2-\frac{J^2\omega^2}{\sin^2\alpha}}\sin2\omega
(t+C)}}.
\end{equation}
We point out that $p_l$ is zero when the extremal values of $|l|$ are reached
given by (4.6).  In the limit $J\to0$ this expression takes the form
\begin{equation}
%<4.11>
p_l = \pm\sqrt{mE}\frac{\sqrt{\frac{l_0^2\omega^2p_{l0}^2}{E^2}}\cos2
\omega t-\left(\frac{l_0^2m\omega^2}{E}-1\right)\sin2\omega t}
{\sqrt{1+\sqrt{\frac{l_0^2\omega^2p_{l0}^2}{E^2}}\sin2
\omega t+\left(\frac{l_0^2m\omega^2}{E}-1\right)\cos2\omega t}}.
\end{equation}
\section{Quantum harmonic oscillator on a cone}
Consider now a quantum particle moving on a double cone and bound to
its tip by the harmonic oscillator potential.  The corresponding
Hamiltonian is
\begin{equation}
%<5.1>
\hat H = \frac{{\hat p_l}^2}{2m} + \frac{{\hat J}^2}{2m{\hat l}^2\sin^2\alpha}
+\frac{m\omega^2}{2}{\hat l}^2.
\end{equation}
Consider the eigenvalue equation for the Hamiltonian
\begin{equation}
%<5.2>
{\hat H}f_E(l,\varphi) = Ef_E(l,\varphi).
\end{equation}
On separating variables and setting
\begin{equation}
%<5.3>
f_{j,E}(l,\varphi) = \frac{1}{\sqrt{|l|}}e^{{\rm i}j\varphi}\tilde u_{j,E}(l),
\end{equation}
we can write (5.2) in the form
\begin{equation}
%<5.4>
\frac{d^2{\tilde u}_{j,E}(l)}{dt^2} + \left(2mE -
\frac{j^2}{l^2\sin^2\alpha}-(m\omega)^2l^2\right){\tilde u}_{j,E}(l) = 0.
\end{equation}
We remark that the potential occuring in the Schr\"odinger equation 
(5.4) which is a sum of the harmonic oscillator and inverse square 
potential is referred to as the Smorodinsky-Winternitz potential 
\cite{12}, isotonic oscillator potential \cite{13} or harmonic 
oscillator with ``centripetal barrier'' \cite{14}.  This potential 
is also closely related to the well-known Morse potential \cite{15}.  
However, as far as we are aware, the Schr\"odinger equation (5.4)
was not discussed in the literature in the context of the double
cone.  The quantum harmonic oscillator in the case of a cone with a
single nappe or equivalently the plane with a deficit angle was
studied in \cite{4} and \cite{6}.  As mentioned earlier this case
is much more complicated than that investigated herein dealing with
quantum mechanics on a double cone, because one is forced to analyse
self-adjoint extensions of symmetric operators representing physical
observables \cite{4,5}.    

Now, making the ansatz
\begin{equation}
%<5.5>
{\tilde u}_{j,E}(l) = C|l|^se^{-\frac{m\omega l^2}{2}}w(m\omega l^2),
\end{equation}
where $C$ is the normalization constant, we find that whenever the
parameter $s$ satisfies
\begin{equation}
%<5.6>
s(s-1) = \frac{j^2}{\sin^2\alpha},
\end{equation}
which leads to
\begin{equation}
%<5.7>
s = \frac{1}{2}\left(1\pm\sqrt{1+\frac{4j^2}{\sin^2\alpha}}\right),
\end{equation}
then $w$ fulfils the Kummer equation
\begin{equation}
%<5.8>
x\frac{d^2w}{dx^2} + \left(s+\frac{1}{2}-x\right)\frac{dw}{dx} - \left(\frac{s}{2}
+\frac{1}{4}-\frac{E}{2\omega}\right)w(x)=0.
\end{equation}
A numerically satisfactory solution to (5.8) with a good behavior at
infinity is given by the Kummer function \cite{16}
\begin{equation}
%<5.9>
w(x) =
U\left(\frac{s}{2}+\frac{1}{4}-\frac{E}{2\omega},s+\frac{1}{2},x\right).
\end{equation}
Another notation for the Kummer function $U(a,b,z)$ is $\Psi(a,b,z)$
\cite{17,18}.  Now, $U(a,b,x)$ is a polynomial when $a=-n$, where
$n=0,\,1,\,2\,\ldots$, and $U(a,b,x)\sim x^{-a}$ for $x\to\infty$.
Demanding decreasing of ${\tilde u}_{j,E}(l)$ as $l\to\infty$ this
leads to the quantization condition on energy
\begin{equation}
%<5.10>
E_{j,n} = 2\omega\left(n+\frac{s}{2}+\frac{1}{4}\right). 
\end{equation}
Furthermore, the requirements $j\ne0$ and convergence of the norm of
the solution to (5.8) rule out the minus sign in (5.7), so
\begin{equation}
%<5.11>
E_{j,n} =
2\omega\left(n+\frac{1}{2}+\frac{1}{4}\sqrt{1+\frac{4j^2}{\sin^2
\alpha}}\right),
\end{equation}
and
\begin{equation}
%<5.12>
w(x) = U\left(-n,1+\frac{1}{2}\sqrt{1+\frac{4j^2}{\sin^2\alpha}},x\right).
\end{equation}
Using the identity \cite{17}
\begin{equation}
%<5.13>
U(-n,\alpha +1,z) = (-1)^nn!L^\alpha _n(z),
\end{equation}
where $L^\alpha_n(x)$ are the generalized Laguerre polynomials, we get
\begin{equation}
%<5.14>
w(x) = (-1)^nn!L^{\frac{1}{2}\sqrt{1+\frac{4j^2}{\sin^2\alpha}}}_n(x).
\end{equation}
Hence, we finally obtain the desired normalized 
(in the sense of $L^2({\mathbb R},dx)$) solution ${\tilde u}_{j,n}(l)$ to 
(5.4) such that
\begin{eqnarray}
%<5.15>
&&{\tilde u}_{j,n}(l)\equiv{\tilde u}_{j,E_n}(l) = (-1)^n
\sqrt{\frac{(m\omega)^{1+\frac{1}{2}\sqrt{1+\frac{4j^2}{\sin^2\alpha}}}n!}
{\Gamma\left(n+1+\frac{1}{2}\sqrt{1+\frac{4j^2}{\sin^2\alpha}}\right)}}
|l|^{\frac{1}{2}\left(1+\sqrt{1+\frac{4j^2}{\sin^2\alpha}}\right)}
e^{-\frac{m\omega l^2}{2}}\nonumber\\
&&\qquad\qquad\qquad\qquad\quad{}\times
L^{\frac{1}{2}\sqrt{1+\frac{4j^2}{\sin^2\alpha}}}_n(m\omega l^2),
\end{eqnarray}
and the normalized solution to the Schr\"odinger equation on the
cone (5.2) of the form
\begin{equation}
%<5.16>
f_{j,n}(l,\varphi) = \frac{1}{2\pi\sqrt{|l|}}e^{{\rm
i}j\varphi}{\tilde u}_{j,n}(l),\qquad j\ne0,
\end{equation}
where the normalization is given by
\begin{equation}
%<5.17>
\langle f_{j,n}|f_{j'n'}\rangle = \int_0^{2\pi}d\varphi\int_{-\infty}^{\infty}dl\,|l|
f_{j,n}^*(l,\varphi)f_{j',n'}(l,\varphi)=\delta_{jj'}\delta_{nn'}.
\end{equation}
Up to the phase factor and normalization constant the solution
(5.15) coincides with the solution to (5.4) introduced by Hakobyan
{\em et al\/} in reference 12.  However, the parameter $k$
labeling the solution obtained in \cite{12} had no physical
interpretation.  On the other hand, it follows immediately from
(5.15) that if $k>\frac{1}{2}$, then this parameter can be 
related to the angular momentum $j$ on the cone via
$k~=~\frac{1}{2}\sqrt{1+\frac{4j^2}{\sin^2\alpha }}$.

Now, an immediate consequence of (5.3) and (5.4) is the following
normalized solution to the eigenvalue equation (5.2) for $j=0$:
\begin{equation}
%<5.18>
f_{0,n}(l) = \frac{1}{2\pi\sqrt{|l|}}\left(\frac{m\omega}{\pi}\right)^\frac{1}{4}
\frac{1}{\sqrt{2^nn!}}e^{-\frac{m\omega l^2}{2}}H_n(\sqrt{m\omega}l),
\end{equation}
referring to the energy
\begin{equation}
%<5.19>
E_{0,n} = \omega(n+\hbox{$\scriptstyle\frac{1}{2}$}),
\end{equation}
where $H_n(z)$ are Hermite polynomials.   Indeed, (5.4) for $j=0$ is
the Schr\"odinger equation for the standard harmonic oscillator on
the line.  Analogously as with the free motion, the solution (5.18)
corresponding to $j=0$ is different from the limit $j\to0$ of the
solution (5.16).  In fact, taking into account the identity \cite{19}
\begin{equation}
%<5.20>
H_{2n+1}(x) = (-1)^n2^{2n+1}n!xL^\frac{1}{2}_n(x^2),
\end{equation}
we obtain the limit $j\to0$ of $f_{j,n}(l)$ given by (5.16) such that
\begin{equation}
%<5.21>
\lim_{j\to0}f_{j,n}(l) = \frac{1}{2\pi\sqrt{|l|}}\left(\frac{m\omega}
{\pi}\right)^\frac{1}{4}\frac{1}{\sqrt{2^{2n+1}(2n+1)!}}
e^{-\frac{m\omega l^2}{2}}H_{2n+1}(\sqrt{m\omega}|l|).
\end{equation}
The energy spectrum in the limit $j\to0$ is
\begin{equation}
%<5.22>
\lim_{j\to0}E_{j,n} = \omega(2n+1+\hbox{$\scriptstyle\frac{1}{2}$}).
\end{equation}
We have thus obtained in the limit $j\to0$ only a part of the
solution corresponding to the case of the harmonic oscillator on a
meridian (generator) related to the odd Hermite polynomials.  As in the case of
the free motion this observation is consistent with instablility of
the solution (4.3) of classical equations of motion describing harmonic
oscillations around a point $l=0$ in the meridian.
\section{Conclusion}
In this work we study the classical and quantum particle on the
double cone in the case of the free motion and harmonic oscillator
potential.  An interesting pecularity of the dynamics on the cone is
the instability of the free motion in the generator on both
classical and quantum level.  This observation seems to be of
importance for better understanding of quantum dynamics.  Indeed, as
with deterministic chaos the role of instabilities in quantum
dynamics is far from clear.   Referring to chaotic motion it is also
worthwhile to point out that the unstable solution describing harmonic
oscillations around a tip of the cone is rather exotic example of
the unstable and bounded solution of a nonlinear dynamical system
which does not show chaotic behavior.  Furthermore, a remarkable
property of the dynamics on the double cone investigated in this
paper is the role played by the vertex which acts as a filter 
selecting from all possible motions the rectilinear one in the 
generator.  Finally, we remark that the results of this paper
provide a physical interpretation as a harmonic oscillator on the
double cone for the one-dimensional Smorodinsky-Winternitz potential
which can be connected with the Morse potential.
\section*{Acknowledgements}
This work was supported by the grant N202 205738 from the National
Centre for Research and Development.

\end{document}